# REFLECTION EQUATIONS AND SURFACE CRITICAL PHENOMENA

M. T. BATCHELOR
*Department of Mathematics, School of Mathematical Sciences, Australian National University, Canberra ACT 0200, Australia*

A brief review is given of recent developments in the study of surface critical phenomena from the viewpoint of exactly solved lattice models. These developments include exact results for the polymer adsorption transition and the surface critical exponents of the eight-vertex model.

## 1 Introduction

There can be little doubt that our understanding of phase transitions and critical phenomena has been greatly enhanced by the study of exactly solved lattice models in statistical mechanics.[1] The early pioneering work of Bethe, Onsager, McGuire, Yang, Lieb and Baxter has left us with the legacy of the Yang-Baxter equation, which governs integrability in the bulk. Many models have been investigated to yield bulk critical behaviour. E.g., Baxter[1,2] derived the bulk free energy of the 8-vertex model, and thus the bulk specific heat

$$C_b \sim |t|^{-\alpha_b} \quad \text{as} \quad t \to 0, \tag{1}$$

where $t = T - T_c$ is a measure of the deviation from criticality and $\alpha_b = 2 - \pi/\mu$ is the continuously varying bulk specific heat exponent, with $\alpha_b = 0(\log)$ in the Ising limit.

In general there is an $R$-matrix $R(u)$ satisfying

$$R_{12}(u)R_{13}(u+v)R_{23}(v) = R_{23}(v)R_{13}(u+v)R_{12}(u) \tag{2}$$

whose elements define the vertex or Boltzmann weights of exactly solvable lattice models in 2-d statistical mechanics. On the other hand, Sklyanin has given a formulation of integrability at a surface, governed by both the Yang-Baxter equation and the reflection equation, which reads[3,4]

$$R_{12}(u-v)K_1^-(u)R_{21}(u+v)K_2^-(v) = K_2^-(v)R_{12}(u+v)K_1^-(u)R_{21}(u-v) \tag{3}$$

where $K^-(u)$ is a reflection or $K$-matrix for the right boundary. A similar equation holds for $K^+(u)$ at the left boundary, which is automatically satisfied

The work reported here has been supported by the Australian Research Council.

if $K^+(u) = K^-(-u-\eta)^t M$, where $\eta$ is a crossing parameter and $M$ is a crossing matrix (both being specific to a given $R$-matrix).[4,5]

Integrable quantum spin chains with open boundaries, such as the XXZ chain,[6] are recovered naturally in the language of $K$-matrices.[4,7] Most of the attention on spin chains has been on the quantum group invariant cases.[7,8] $K$-matrices have been found for a number of models, although up until quite recently very little in the way of surface critical phenomena has been derived from them.

The most natural open geometry for vertex models is that depicted in Fig. 1. The integrable boundary weights follow directly from the $K$-matrix elements. Specifically, the left and right boundary weights are given by [9]

$$l_a^b(u) = \sum_{ab} R_{ac}^{db}(2u) K_{cd}^+(u), \quad r_a^b(u) = K_{ab}^-(u). \tag{4}$$

I give a brief account in Sec. 2 of results obtained with Ming Yung for the dilute O($n$) loop model with open boundaries and related self-avoiding polymer problems.[8–12] I then discuss more recent work [14] with Yu-Kui Zhou in Sec. 3 on surface critical phenomena in the 8-vertex model.[a] Further developments are mentioned in Sec. 4.

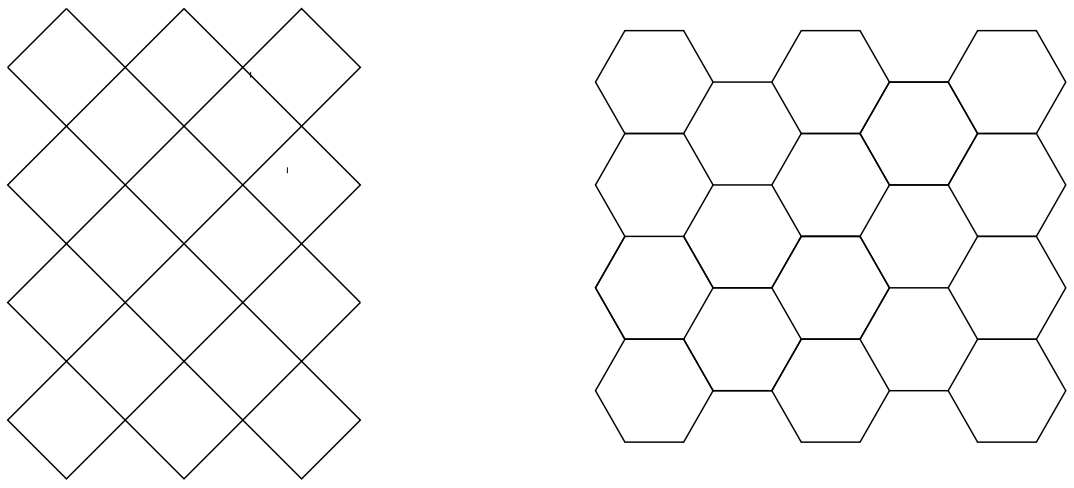

Figure 1: The open square and honeycomb lattices.

## 2 The Dilute O($n$) Loop Model and Related Polymer Problems

The dilute O($n$) loop model on the square lattice [15] turns out to be solvable for two sets of boundary weights.[9] The partition function is defined as

$$Z_{\text{loop}} = \sum_{\mathcal{G}} \rho_1^{m_1} \cdots \rho_{13}^{m_{13}} \, n^P \tag{5}$$

[a] Our results for the 8-vertex model were announced at this Meeting.

where the sum is over all configurations $\mathcal{G}$ of non-intersecting closed loops which cover some (or none) of the edges of the square lattice in Fig. 1. The possible vertex configurations are shown in Fig. 2, with a vertex of type $i$ carrying a Boltzmann weight $\rho_i$. In configuration $\mathcal{G}$, $m_i$ is the number of occurrences of the vertex of type $i$, while $P$ is the total number of closed loops of fugacity $n$.

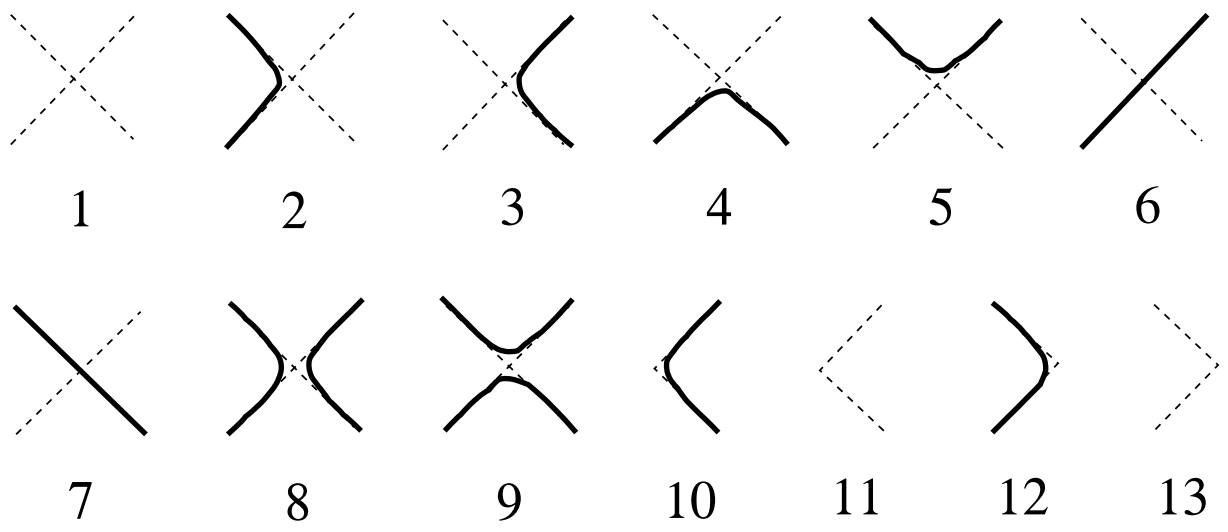

Figure 2: The allowed loop model vertices and corresponding weights.

The integrable bulk loop weights are [15,16]

$$\begin{aligned}
\rho_1 &= \sin(3\lambda - u)\sin(u) + \sin(2\lambda)\sin(3\lambda) \\
\rho_2 = \rho_3 &= \epsilon_1 \sin(3\lambda - u)\sin(2\lambda) \\
\rho_4 = \rho_5 &= \epsilon_2 \sin(u)\sin(2\lambda) \\
\rho_6 = \rho_7 &= \sin(3\lambda - u)\sin(u) \\
\rho_8 &= \sin(3\lambda - u)\sin(2\lambda - u) \\
\rho_9 &= -\sin(\lambda - u)\sin(u).
\end{aligned} \tag{6}$$

where $\epsilon_1^2 = \epsilon_2^2 = 1$ and $n = -2\cos(4\lambda)$. The two sets of integrable boundary loop weights are [9]

$$\text{(A)} \quad \begin{aligned} \rho_{10} = \rho_{13} &= \sin[\tfrac{1}{2}(3\lambda - u)] \\ \rho_{11} = \rho_{12} &= \epsilon_1 \sin[\tfrac{1}{2}(3\lambda + u)] \end{aligned} \tag{7}$$

$$\text{(B)} \quad \begin{aligned} \rho_{10} = \rho_{13} &= \cos[\tfrac{1}{2}(3\lambda - u)] \\ \rho_{11} = \rho_{12} &= \epsilon_1 \cos[\tfrac{1}{2}(3\lambda + u)]. \end{aligned} \tag{8}$$

These weights follow from the $K$-matrices of the underlying Izergin-Korepin vertex model.[8] The eigenvalues of the transfer matrix have been determined for both cases by means of the Bethe Ansatz.[9]

Exact results have been obtained for various polymer problems on the related honeycomb lattice, which follows in the honeycomb limit [15] $u = \lambda$ (where the bulk weight $\rho_9 = 0$). In this way loop configurations on the square lattice become loop configurations on the honeycomb lattice depicted in Fig. 1, where the partition function reduces to [10]

$$Z_{\text{loop}} = \sum_{\mathcal{G}} x^L y^{L_s} \, n^P. \tag{9}$$

At $n = 0$ the contribution from all closed loops vanishes and the partition sum reduces to a generating function for self-avoiding walks where $x$ is the fugacity of a step in the bulk and $y$ is the fugacity of a step along the surface, in this case either edge of the strip. Here $L$ is the length of the walk in the bulk and $L_s$ is the length of a walk along the surface.

The values of $x$ and $y$ are fixed by integrability. The well known Nienhuis result [17]

$$1/x^* = \sqrt{2 \pm \sqrt{2-n}} \tag{10}$$

follows from the bulk weights. On the other hand, at the boundary $y^* = x^*$ for case (A), i.e., the critical surface coupling is equal to the bulk coupling. In the language of surface critical phemonena,[18] this corresponds to an integrable point on the *ordinary* transition line. For case (B)

$$1/y^* = \sqrt{\pm\sqrt{2-n}}. \tag{11}$$

This point corresponds to the $\mathrm{O}(n)$ model at the *special* transition.

The polymer adsorption transition corresponds to the $n = 0$ limit of the $\mathrm{O}(n)$ model at the *special* surface transition.[18] At $n = 0$ we have

$$1/x^* = \sqrt{2+\sqrt{2}} \quad \text{and} \quad 1/y^* = \sqrt{\sqrt{2}}. \tag{12}$$

The exact critical adsorption temperature $T_a$ is thus given by [10,11]

$$\exp\left(\frac{\epsilon}{kT_a}\right) = y^*/x^* = \sqrt{1+\sqrt{2}} = 1.553\ldots \tag{13}$$

where $\epsilon$ is the contact energy at the surface. Physically, there is a desorbed phase for $T > T_a$ and an adsorbed phase for $T < T_a$. The number of configurations of self-avoiding walks with one end attached to the boundary has the asymptotic form [19]

$$Z_1 \sim \begin{cases} \mu^L L^{\gamma_1^{\text{o}}-1} & T > T_a \\ \mu^L L^{\gamma_1^{\text{sp}}-1} & T = T_a \end{cases} \tag{14}$$

where for the honeycomb lattice, $\mu = 1/x^*$. The surface critical exponents $\gamma_1^{\rm o}$ and $\gamma_1^{\rm sp}$ can be obtained from the finite-size corrections to the transfer matrix eigenvalues, with result [10,13,20]

$$\gamma_1^{\rm o} = \tfrac{61}{64} \quad \text{and} \quad \gamma_1^{\rm sp} = \tfrac{93}{64}. \tag{15}$$

The value for $\gamma_1^{\rm o}$ is the conformal invariance result of Cardy,[21] while the value for $\gamma_1^{\rm sp}$ had originally been conjectured by Guim and Burkhardt.[19] The exact value of the crossover exponent also follows as $\phi = \frac{1}{2}$.

A further configurational exponent,

$$\gamma_1 = \tfrac{85}{64}, \tag{16}$$

is obtained by considering *mixed* ordinary and special boundary conditions arising from different $K$-matrices on either side of the strip.[12,13] This case can also be solved exactly by means of the Bethe Ansatz and the exponent $\gamma_1$ is seen to control the number of self-avoiding configurations beginning from near some origin at the surface with a non-adsorbing boundary to the left and an adsorbing boundary to the right (see Fig. 3). The same adsorption temperature $T_a$ holds for the mixed boundary case. These results have been numerically confirmed in the half-plane using series expansion techniques.[22]

Although we have obtained various exact exponents by explicit calculations on the honeycomb lattice, they are expected to be valid for other planar lattices on universality grounds.

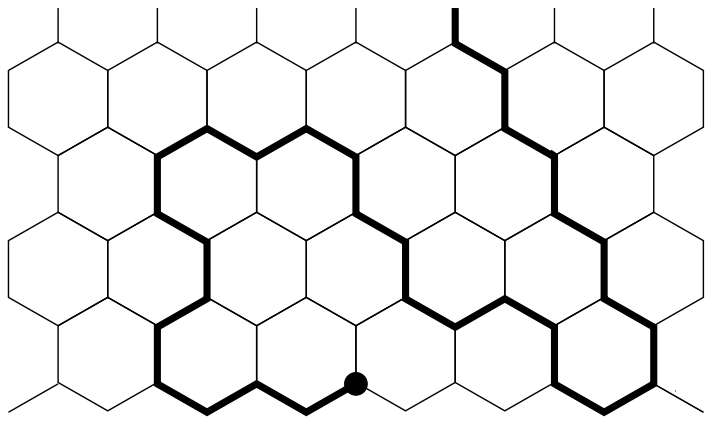

Figure 3: A self-avoiding walk from an origin on the surface of the half-plane.

## 3 The 8-vertex Model

$K$-matrices have also been found for the 8-vertex model.[23] As for bulk weights,[24,25] the boundary weights can be interpreted in terms of Ising interactions (see Fig. 4). The parametrisation of the vertical and horizontal Ising couplings $K$ and

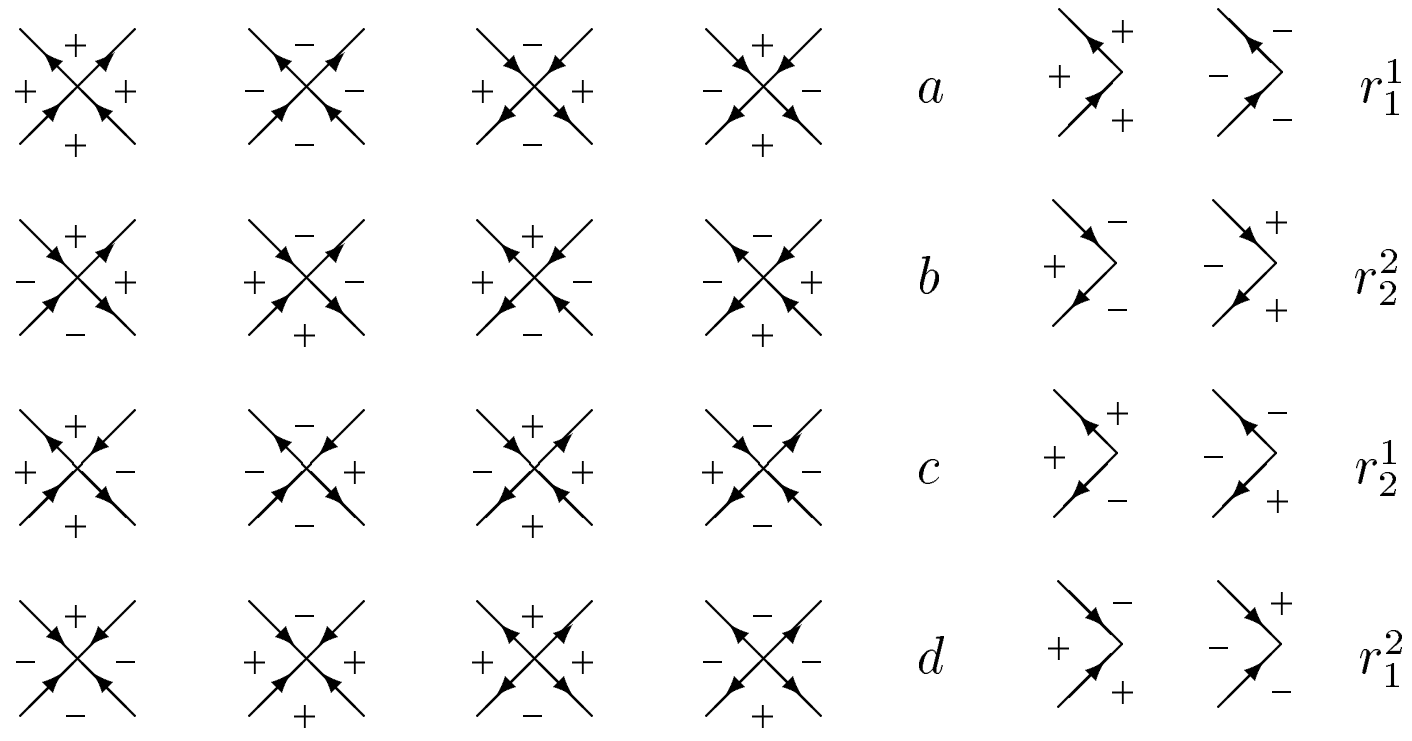

Figure 4: The bulk and surface vertex and Ising spin configurations.

$L$ and the four-spin interaction $M$ is given in terms of elliptic theta functions as [24,25,1]

$$e^{4K}=\frac{ac}{bd}=\left[\frac{\theta_4(u)}{\theta_1(u)}\right]^2,\quad e^{4L}=\frac{ad}{bc}=\left[\frac{\theta_1(\lambda-u)}{\theta_4(\lambda-u)}\right]^2,\quad e^{4M}=\frac{ab}{cd}=\left[\frac{\theta_4(\lambda)}{\theta_1(\lambda)}\right]^2.$$

Recalling (4), the Ising coupling $K_s$ in the surface layer follows in terms of the known $K$-matrix as

$$e^{4K_s}=\frac{r_1^1r_2^2}{r_1^2r_2^1}=\frac{K_{11}^-(u/2)K_{22}^-(u/2)}{K_{12}^-(u/2)K_{21}^-(u/2)}. \tag{17}$$

In this way we can realize an anisotropic Ising model with four-spin interactions with open boundary conditions (see Fig. 5). Although we have the integrable boundary weights, so far there is no exact solution for this model. However, an inversion relation for the free energy has been obtained by Yu-Kui Zhou.[26] We have used the inversion relation method [1] to solve this relation for the surface free energy and thus the surface specific heats [18] $C_s$ and $C_1$. We find that as $t\to 0$,

$$C_s\sim|t|^{-\alpha_s}\qquad\text{with}\qquad \alpha_s=2-\frac{\pi}{2\mu}, \tag{18}$$

$$C_1\sim|t|^{-\alpha_1}\qquad\text{with}\qquad \alpha_1=1-\frac{\pi}{\mu}. \tag{19}$$

The known Ising results,[18] $\alpha_s=1(\log)$ and $\alpha_1=-1(\log)$, are recovered at $\mu=\pi/2$. Our results are in agreement with the scaling relations $\alpha_s=\alpha_b+\nu$ and $\alpha_1=\alpha_b-1$,[18] where $\nu=\pi/2\mu$ is the correlation length exponent.[27]

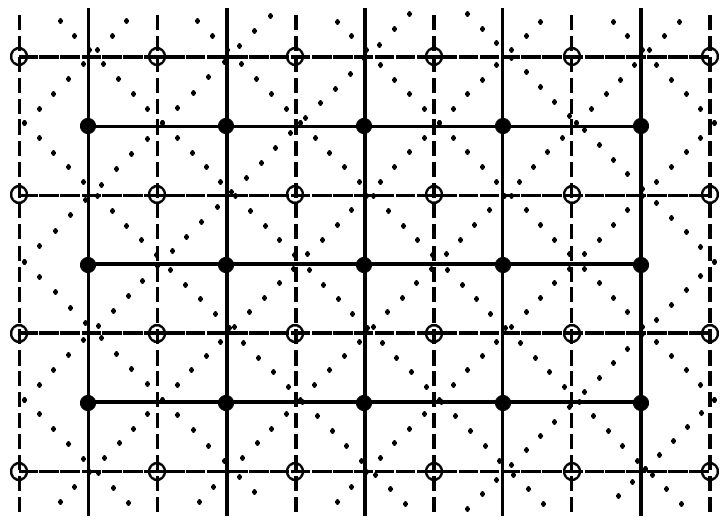

Figure 5: The 8-vertex lattice (dotted lines) and the Ising lattice (broken and solid lines).

## 4 More Recent Developments

Sklyanin's approach can be reformulated in the language of face models in which the $K$-matrices are interpreted as boundary face weights of a double row transfer matrix. This has recently been done by a number of authors.[26,28–31] Such $K$-matrices were presented at this Meeting for the Andrews-Baxter-Forrester model in the talks by Koo and Pearce.[30,32] There have been some rapid developments in this area since this Meeting. First, we repeated our 8-vertex model calculations for the Andrews-Baxter-Forrester model to derive the excess specific heat exponent $\alpha_s = (7-L)/4$, with the Ising value again recovered for $L = 3$.[33] We also derived the exponents $\alpha_s$ and $\alpha_1$ for the CSOS model [34] and the exponents $\delta_s$ and $\alpha_s$ for the dilute $A_L$ models. This latter work yielded the value $\delta_s = -\frac{15}{7}$ for an Ising model in a magnetic field.[35] Further $K$-matrices have been found for general face and vertex models associated with the algebras $A_n^{(1)}, B_n^{(1)}, C_n^{(1)}, D_n^{(1)}$ and $A_n^{(2)}$.[36]